\documentclass[prl,twocolumn,showpacs]{revtex4}
\usepackage{bm}
\usepackage{graphicx}
\usepackage{amssymb}
\usepackage{amsmath}
\usepackage{ulem}
\usepackage{color}
\newcommand{\nix}[1]{}

\begin{document}
%
%
\title{Terahertz radiation driven chiral edge currents in graphene}
\author
{J.~Karch,$^{1}$  C.~Drexler,$^1$ P.~Olbrich,$^1$
  M.~Fehrenbacher,$^{1}$ M.~Hirmer,$^1$ M.~M.~Glazov,$^2$
S.~A.~Tarasenko,$^2$ E.~L.~Ivchenko,$^2$ B.~Birkner,$^1$ J.~Eroms,$^1$
D.~Weiss,$^1$
R.~Yakimova,$^3$ S.~Lara-Avila,$^4$ S.~Kubatkin,$^4$
%
M.~Ostler,$^{5}$
T.~Seyller,$^{5}$ and
S.~D.~Ganichev$^{1}$}

\affiliation{$^1$ Terahertz Center, University of Regensburg,
93040 Regensburg, Germany}
\affiliation{$^2$ Ioffe Physical-Technical Institute, Russian Academy
  of Sciences,
194021 St.~Petersburg, Russia}
\affiliation{$^3$
Link{\"o}ping University,
S-58183 Link{\"o}ping, Sweden}
\affiliation{$^4$
Chalmers University of Technology,
S-41296 G{\"o}teborg, Sweden}
\affiliation{$^5$
University of Erlangen-N{\"u}rnberg,
91058 Erlangen, Germany}

\begin{abstract}
We observe photocurrents induced in
single layer graphene samples
by  illumination of the graphene edges with circularly polarized
terahertz radiation at normal incidence.
The photocurrent flows along the sample edges and forms a vortex.
Its winding direction reverses by switching the light helicity from
left- to right-handed. We demonstrate that the photocurrent stems from the sample edges, which reduce
the spatial symmetry and result in an asymmetric scattering of carriers driven by the radiation
electric field. The developed theory is in a good agreement with the experiment. We show that the
edge photocurrents can be applied for determination of
the conductivity type and the momentum scattering time of the charge carriers in the
graphene edge vicinity.
\end{abstract}



\date{\today}

\maketitle

The ``bulk'' transport properties of graphene have been studied intensively in
recent years and yielded insight into the half-integer and fractional
quantum Hall effect, phase-coherent effects or spin transport on the micrometer
scale, to name a few examples~\cite{Bib:Novoselov2004,sarma}.
While the details of each of those effects
depend crucially on the linear dispersion relation of graphene and its specific material
properties, most of the transport phenomena have already been studied in other
two-dimensional systems.
Graphene edges, on the other hand, were predicted to show insulating or metallic, even magnetic
behavior, depending on the crystallographic orientation and edge chemistry.
In scanning tunneling experiments, an enhanced edge
density of states was shown~\cite{Tun2,Tun4} and
Raman scattering
experiments provided evidence for
the dependence of scattering mechanisms on the edge orientation~\cite{Raman1,Raman3}.
In transport experiments edge effects are usually masked by bulk properties, 
nonetheless the graphene edges are expected to play a crucial role in the electronic 
properties of graphene-based nanoscale devices.

\begin{figure}[b]
\includegraphics[width=0.85\linewidth]{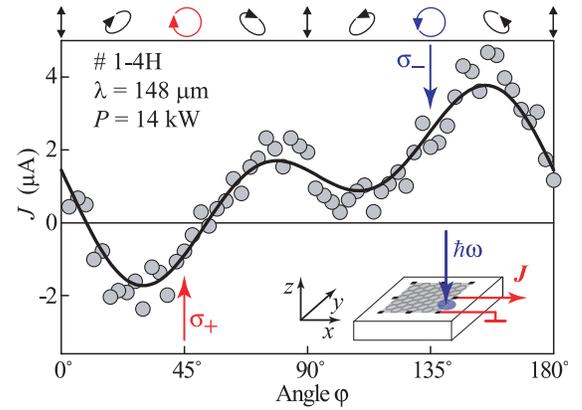}
\caption{
Photocurrent in sample \#1-4H  as a function of the
angle $\varphi$ defining the light polarization.
Solid line is a fit to  Eq.~\protect(\ref{phenom1}) [see
also Eq.~\protect(\ref{J_y_3}) and discussion].
%
The inset shows the experimental geometry.
The ellipses on top illustrate the polarization states for various
$\varphi$.
} \label{figure1}
\end{figure}

Here, we present an opto-electronic method to uniquely distinguish edge from
bulk scattering by exploring edge photocurrents in graphene samples illuminated
by terahertz (THz) radiation. For circularly polarized light 
the edge current is observed to form a vortex winding around the edges of the square-shaped 
samples. Its direction reverses upon switching the radiation helicity from 
left- to right-handed. Evidently, the photocurrent is caused by the local 
symmetry breaking at the sample edges resulting in an asymmetric scattering 
of carriers driven by the radiation electric field.
It gives rise to a directed electric current along the sample boundary in a
narrow stripe of width comparable to the mean free path.
We show that the photocurrent  measurements provide direct access to electron 
scattering at the graphene edges and allow  to 
map the variation of scattering times along the edges.

\begin{figure}[t]
\includegraphics[width=0.65\linewidth]{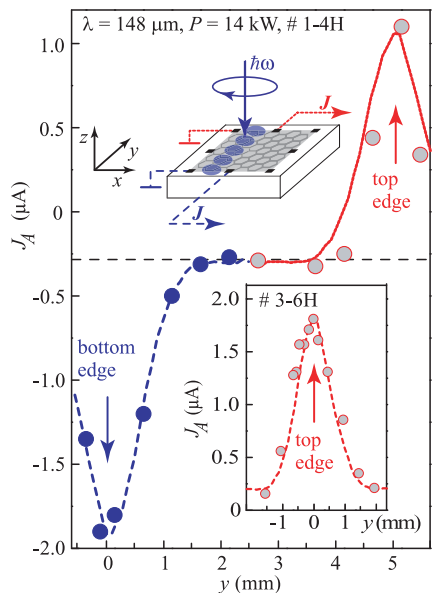}
\caption{Photocurrent $J_{A}$ in sample~\#1-4H  as a function of the laser spot position.
The laser spot is scanned along $y$ and the current is
picked up from two contact pairs at the top
(red circles) or bottom (blue full circles) sample edges aligned along $x$  (see inset).
Dashed lines represent the laser beam spatial distribution, which is measured by a
pyroelectric camera, scaled to the current maximum.
The bottom inset shows the scan for sample~\#3-6H.
} \label{figure2}
\end{figure}

\begin{figure}[t]
\includegraphics[width=0.9\linewidth]{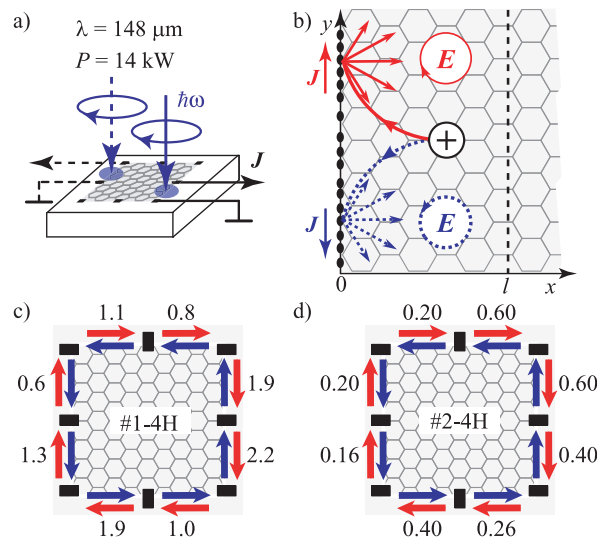}
\caption{ (a) Experimental geometry for the study of edge photocurrents.
(b) Schematic illustration of the edge current generation.
The electric field of circularly polarized radiation rotates clockwise or counterclockwise
resulting in a circular motion of carriers, which is sketched by red and blue trajectories, respectively.
Our theoretical model, see Eq.~(\ref{J_y_3}), shows that the circular edge current stems from carriers moving towards the edge.
It is due to the \textit{second order}  \textbf{\textit{E}}-field correction
to the distribution function and involves the retardation of the electron motion with respect to the instantaneous electric field.
Switching the radiation helicity reverses the motion direction and, consequently, the electric current. (c) and (d) photocurrent topology.
Red and blue arrows show the current direction for $\sigma_+$ and $\sigma_-$
polarizations, respectively. Numbers indicate the photocurrent amplitude
$J_{A}$ in microampers.
}
\label{figure3}
\end{figure}

We investigated two types of single-layer graphene samples: (i) large-area epitaxial 
graphene prepared by high-temperature 
Si sublimation of 4H and 6H polytypes of semi-insulating SiC substrates~\cite{erl1,LaraAvival09,erl2} 
and (ii) small
area exfoliated graphene flakes~\cite{Bib:Novoselov2004}
deposited on oxidized silicon 
wafers.
Below, we report results on epitaxial graphene
samples (labeled \#1-4H, \#2-4H, and \#3-6H) and three 
samples prepared from exfoliated graphene.
Hall measurements indicate that the epitaxial
samples are $n$-doped (due to
charge transfer from SiC~\cite{erl1}) 
while the exfoliated samples are $p$-doped.
The measured carrier density lies in the range
$( 2 \div 7 ) \times$10$^{12}$ cm$^{-2}$,
the Fermi energy $E_F$ ranges from  200 to 300~meV and the mobility is about 1000\,cm$^2$/Vs at
room temperature.
Ohmic contacts were made
at samples' edges (see, e.g., inset of Fig.~\ref{figure1}).
%
%
Details on the material growth and characterization
can be found in~\cite{Suppl}.

The experiments on edge photocurrents are performed applying
alternating electric THz fields of a high power pulsed NH$_3$ laser~\cite{JETP1982,laser2,Ganichevbook}
operating at  wavelengths $\lambda \, = \, 90.5\,\mu$m, $148\,\mu$m or 280\,$\mu$m
(frequencies $f \, = \, 3.3$\,THz, 2\,THz and 1.1\,THz, respectively).
The radiation induces indirect (Drude-like) optical
transitions, because the photon energies are much smaller than the carrier Fermi
energy.
The  NH$_3$ laser generates single pulses with a duration
of about 100~ns, peak power of $P \approx $ 10~kW, and a
repetition rate of 1~Hz.
A typical spot diameter
from 1 to 3~mm. The beam has an almost Gaussian form,
which is measured by a
pyroelectric camera~\cite{ch1Ziemann2000p3843}.

All experiments are performed at normal incidence of light and at
room temperature. Elliptically and, in particular, circularly polarized radiation
is obtained applying
$\lambda$/4
quartz plates. The resulting polarization state described by
the Stokes parameters~\cite{Stokes} $S_1$, $S_2$, and $S_3 \equiv P_{\rm circ}$
is directly related to the angle $\varphi$ between the initial linear
polarization of the laser light along the $y$-axis and the plate optical axis.
%
%
The experimental geometry is 
shown
in Figs.~\ref{figure1}, \ref{figure2}, and~\ref{figure3}.
The 
current
is measured via the voltage drop  across a 50\,$\Omega$ load resistor.

Illumination of the edge of unbiased large-area samples between any pair of contacts
results in a photocurrent. By contrast, if the laser spot is moved toward the 
center the signal vanishes. The detected signal depends strongly on the radiation polarization, 
Fig.~\ref{figure1}.
The principal observation is that for right- ($\sigma_+$) and left-handed ($\sigma_-$) polarizations,
i.e., for $\varphi = 45^{\circ}$ and 135$^{\circ}$, the signs of the photocurrent $J$ are
opposite. The overall dependence $J(\varphi)$ is more complex. It is well described by
\begin{eqnarray}
J(\varphi) &=& J_{A} \sin 2\varphi + (J_{B}/2) \sin 4\varphi - J_{C}\cos^2 2\varphi +  \xi
\nonumber \\
&=& J_A P_{\rm{circ}}(\varphi) + J_B S_2(\varphi) + J_C   S_1(\varphi)+ \xi \label{phenom1}
\end{eqnarray}
and
corresponds to the superposition of the Stokes parameters with different weights.
%
The first term given by the coefficient $J_{A}$ is just proportional to the
radiation helicity, whereas the second ($J \propto J_{B}$) and third ($J \propto J_{C}$) terms
change with degree and orientation of the linear polarization.
Note that the observed offset $\xi$
is usually smaller or comparable to $J_{A}$, $J_{B}$, and  $J_{C}$ (see Fig.~\ref{figure1}).
In our present study we focus on the helicity driven photocurrent $J_{A}$.
This is the only contribution which reverses the current direction upon switching the radiation
helicity from $\sigma_+$ to $\sigma_-$.
We also note that, for circularly polarized light ($P_{\rm circ} = \pm 1$ and $S_1 = S_2 = 0$) and $\xi = 0$,
the current is solely determined by the first term in Eq.~(\ref{phenom1}).
Therefore, it would be sufficient to measure the response to circularly polarized radiation.
However, to increase the accuracy,
we always measured the whole polarization
dependence, like the one shown in Fig.~\ref{figure1},
and extracted
$J_{A}$ by fitting  Eq.~(\ref{phenom1}) to the data.
%
%

\begin{figure}[t]
\includegraphics[width=0.85\linewidth]{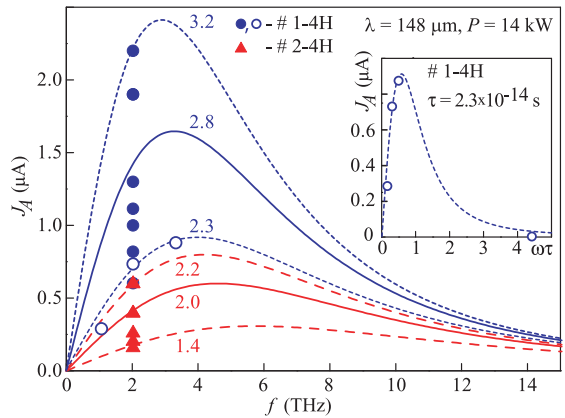}
\caption{Photocurrent $J_A$ measured at 2~THz for different edge segments 
[see Figs.~\protect\ref{figure3}c and \protect\ref{figure3}d for circles and triangles, respectively]. 
Lines are fits 
to  Eq.~\protect(\ref{J_y_3}). 
The 
fitting parameters $\tau / 10^{-14}$\,s for 
sample~\#1-4H and \#2-4H are indicated by numbers. 
The inset shows the measured circular photocurrent $J_A(\omega \tau)$ at one of the edge segments of sample~\#1-4H
(open circles) together with the fit after Eq.~\protect(\ref{J_y_3}).
Data point at $\omega \tau = 4.4$ is obtained applying pulsed CO$_2$ laser.
}
\label{figure5}
\end{figure}

To prove that the photocurrent is caused by illuminating the graphene edges, we
scanned the laser spot across the sample along the $y$-axis.
The signal was picked up from a pair of contacts at
the sample top and bottom edges aligned along the $x$-axis.
The experimental geometry and the photocurrent $J_{A}$ versus the
spot position are shown in Fig.~\ref{figure2}.
The current reaches its maximum for the
laser spot centered at the edge
and rapidly decays with the spot moving.
Comparison of $J_{A}(y)$ with the independently recorded
laser profile (dotted line) shows that
the signal just follows the Gaussian intensity profile.
%
This observation unambiguously demonstrates that the
photocurrent is caused by illuminating the sample edges.
Moreover, Fig.~\ref{figure2} reveals that the helicity driven current $J_{A}$
changes its sign for opposite edges.

The above results show
that the current direction at a specific edge depends on the light helicity.
To check this in more detail, we investigated currents excited by circularly
polarized radiation for different pairs of contacts. Here, the
laser spot is  always
centered between the contacts, see Fig.~\ref{figure3}a.
The current direction for $\sigma_+$ (red arrows) and $\sigma_-$ (blue arrows) circularly polarized radiation
and the magnitude of $J_{A}$ for various contact pairs are shown in Fig.~\ref{figure3}b and~\ref{figure3}c.
The figures document a remarkable behavior of the circular
edge photocurrent: it forms a vortex winding around the edges of the square shaped samples
which reverses its direction upon switching from
right- to left-handed.
These dependencies are observed for
all used wavelengths and samples. Helicity driven currents have also
been observed for  small area graphene flakes, see~\cite{Suppl} for details.

%
The observation that a  photocurrent
occurs only if the laser spot is adjusted to an edge agrees with
the symmetry analysis: In the ideal honeycomb lattice of graphene
and for our experimental geometry,
any photoelectric effect is forbidden~\cite{PRL10},  because the
two-dimensional structure of graphene possesses a center of space
inversion.
Thus, the appearance of photocurrents at \textit{normal} incidence of radiation is a clear
manifestation of the symmetry reduction of the system, in our case, due to the edges.
We also note that the typical photon energy $\hbar\omega$$\sim$10~meV used in  experiment
is much smaller than the characteristic energy of carriers $E_F \sim 100$~meV.
%
Thus, the mechanism of  current formation can be treated classically
and should involve the action of
the light's electric field on free carriers in the vicinity of a graphene edge.

A microscopic process actuating the edge photocurrent generation is
illustrated in Fig.~\ref{figure3}b. It involves the time dependent
motion of carriers  under the action of the electric field of circularly
polarized radiation and scattering at the sample edge.
We note that this mechanism is  similar to that of the surface photogalvanic effect
observed in bulk materials~\cite{Magarill,Alperovich}.
%
%
The microscopic theory of edge currents is developed in the framework of
the Boltzmann kinetic equation.
In this approach, the electron (hole) distribution is described by the
function $f(\bm{p},x,t)$.
It depends on the carrier momentum $\bm p$, coordinate $x$
($x \geq 0$ for a semi-infinite layer), time $t$, and obeys the equation
\begin{equation}\label{f_general}
\frac{\partial f}{\partial t} + v_x  \frac{\partial f}{\partial x}
+ q \bm{E}(t) \frac{\partial f}{\partial \bm{p}} = Q\{f\} \:,
\end{equation}
where $\bm{E}(t) = \bm{E}_0 {\rm e}^{- \mathrm i \omega t} + \bm{E}_0^* {\rm e}^{+\mathrm  i \omega t}$
is the electric field of the radiation,
$\bm{v} = v \bm p/p$ is the electron velocity,
$v \approx 10^6$~m/s is the effective speed, $q$ is the carrier charge ($q= + |e|$ for holes and $- |e|$ for
electrons), and $Q\{f\}$ is the collision integral.
The distribution function can be expanded in series of powers of the electric field,
\begin{equation}
\label{f:gen}
f(\bm{p},x,t) = f_0(\varepsilon_{\bm{p}}) + [f_1 (\bm{p},x){\rm e}^{- \mathrm i \omega t} + {\rm c.c.}]
+ f_2(\bm{p},x) + ... \:,
\end{equation}
where $f_0(\varepsilon_{\bm{p}})$ is the equilibrium distribution
function with $\varepsilon_{\bm{p}} = vp$ being the electron energy,
$f_1 \propto |\bm{E}|$, and $f_2 \propto |\bm{E}|^2$.
The first order in $\bm E$ correction to the distribution function
oscillates with frequency $\omega$ and does not contribute to a $dc$ current.
The directed electric current along the structure edge is, therefore,
determined by the \textit{second order}  \textbf{\textit{E}}-field correction $f_2$ and
given by
\begin{equation}\label{J_edge_gen}
J_y = 4\ q \int_{0}^{\infty} dx \sum_{\bm{p}} f_2(\bm{p},x) v_y \:.
\end{equation}
The factor~4 accounts for the spin and valley degeneracy.

We solve Eq.~\eqref{f_general} and calculate the current, Eq.~\eqref{J_edge_gen},
for the simple form of the collision integral,
\begin{equation}
Q \{ f(\bm{p},x,t) \} = - \frac{f(\bm{p},x,t) - f_0(\varepsilon_{\bm{p}})}{\tau},
\end{equation}
with $\tau$ being the scattering time, and the boundary condition at $x=0$,
\begin{multline}
\label{bound}
 f(p_x > 0,p_y,0,t) = \\
 - \int   \frac{v_x'}{2p'} f(\bm p',0,t) \delta(\varepsilon_{\bm p}-\varepsilon_{\bm p'}) \Theta(-v_x')d\bm p'\,,
\end{multline}
corresponding to diffusive scattering.
%
In the case of
a degenerate gas,
the
edge current takes the form (see~\cite{Suppl} 
for details):
\begin{multline}\label{J_y_3}
J_y = - \frac{q^3 \tau^3 v^2}{2\pi \hbar^2  [1+(\omega\tau)^2]}
\left[\frac{10}{3} \, \frac{\omega\tau}{1+(\omega\tau)^2} {\rm i}[\bm{E}_0 \times \bm{E}_0^*]_z + \right. \\
\left. \left( 1 + \frac{7}{6} \, \frac{1-(\omega\tau)^2}{1+(\omega\tau)^2} \right) \times (E_{0,x} E_{0,y}^* + E_{0,y} E_{0,x}^*)
 \right]
\:.
\end{multline}
%
The helicity-driven 
current is given by the first term
because ${\rm i}[\bm{E}_0 \times \bm{E}_0^*]_z \equiv - P_{\rm{circ}}$ for our geometry where the  light propagates along $-z$.
The second term yields the
current caused by linearly polarized radiation and vanishes for circular polarization.
In the case of elliptically polarized light,
$E_{0,x} E_{0,y}^* + E_{0,y} E_{0,x}^* \propto (1/2) \sin{4\varphi} = S_2$.
Both contributions are clearly detected in the experiment and correspond to the
first ($\propto J_{A}$) and  second ($\propto J_{B}$)
terms in the empirical Eq.~(\ref{phenom1}), see Fig.~\ref{figure1}.

The helicity driven
photocurrent described by
Eq.~(\ref{J_y_3}) vanishes for zero frequency, has a maximum at
$\omega\tau \simeq 0.6$ and decreases rapidly at higher frequencies.
Exactly this behavior is found in experiment (see inset of
Fig.~\ref{figure5}) as we explain in more detail below. The only free
parameter in Eq.~\eqref{J_y_3} is the scattering time $\tau$.  
Corresponding data are shown in Fig.~\ref{figure3} where the
photocurrent values measured at 2~THz for each of the contact pairs
are plotted. These data points are first compared to calculated traces of
$J_{A}$ employing Eq.~\eqref{J_y_3}. Solid lines are calculated using
the bulk values for the 
time $\tau$ extracted from resistivity and carrier density for samples~\#1-4H
($\tau=2.0 \times 10^{-14}$~s) and~\#2-4H ($\tau=2.8 \times 10^{-14}$~s).
The bulk scattering times used in Eq.~\eqref{J_y_3} give for some 
of the contact pairs already perfect quantitative agreement.
For other edge segments the current deviates significantly. This 
is a consequence of the strongly non-linear dependence of $J_{A}$ on $\tau$. 
Varying $\tau$  by only $\pm$15\% changes the current by $\pm 50$~\%. By
fitting the photocurrent $J_A$ we can extract the local scattering time
$\tau$ for every edge segment shown in Figs.~\ref{figure3}c and
d). The best fits are shown by dashed lines in Fig.~\ref{figure5} and constitute 
a map of scattering times along the edge. The average value of the circular edge 
current scales with the sample mobility. To check the frequency dependence predicted 
by Eq.~\eqref{J_y_3} we show in the inset of Fig.~\ref{figure5} $J_{A}$ vs. 
$\omega \tau$ for one edge segment using the extracted $\tau$. The data points are perfectly 
described by Eq.~\eqref{J_y_3} and confirm the model. 

While the magnitude of the circular edge photocurrent agrees well with  theory,
the expected polarity of $J_A$ for $n$-type graphene is opposite to the one observed.
%
This, at  first glance, surprising result 
agrees with results from spatially resolved Raman measurements 
demonstrating that edges of $n$-type graphene layers
exhibit $p$-type conductivity~\cite{Raman1,Raman3}.
This explains the sign of the photocurrent, which is generated
in a narrow edge channel comparable to the mean free path ($\approx 10 \div 20$~nm) and
has opposite sign for electrons and holes, see Eq.~(\ref{J_y_3}).
%
Actually, the difference in the conductivity type
can be also understood from the details of the sample fabrication.
It is well established  that epitaxial
graphene on SiC(0001) is $n$-doped due to charge transfer
from the interfacial buffer layer (see, e.g.,~\cite{erl1,LaraAvival09}),
while so-called quasi-free-standing graphene,
lacking such buffer layer and sitting  on a hydrogen terminated SiC(0001) surface, is
$p$-doped~\cite{erl5}. Therefore, it is reasonable that the edges of epitaxial graphene,
exposed to the SiC substrate without the interfacial layer, can  be $p$-doped. This
assumption is corroborated by similar reports on the transition from $n$-
to $p$-type of doping at the edges of graphene flakes on SiO$_2$, which
were attributed to the difference in the work functions of
graphene and the substrate~\cite{SK_TS4}.


To summarize, our observations clearly demonstrate that 
illuminating monolayer graphene edges with polarized terahertz radiation
at normal incidence results in a directed electric edge current.
The effect is directly coupled to  electron scattering
at the graphene edge  and vanishes in bulk graphene.
Our results  suggest that circular the photocurrents
can be effectively used to study edge transport in graphene 
even at room temperature.

We thank   K.~S. Novoselov, V. Lechner, S. Heydrich and V.~V. Bel'kov
for fruitful discussions.
Support from DFG (SPP~1459 and GRK~1570), EU-ConceptGraphene, Linkage Grant of IB of BMBF at DLR, RFBR,
Russian Ministry of Education and Sciences, and ``Dynasty'' Foundation--ICFPM
is acknowledged.

\newpage

\section{Supplemental Material}

\subsection{S1. Details of the Samples}
\label{samples}

We investigated three epitaxial samples grown on SiC. Samples~\#1-4H and
\#2-4H were grown by the Link{\"o}ping group on a Si-terminated
surface of a 4H-SiC(0001) semi-insulating substrate (Cree
Inc.)~\cite{LaraAvival09}. The reaction kinetics on the
Si-terminated surface is slower than on the C-face because of the
higher surface energy, which fosters homogeneous and well controlled
graphene formation~\cite{erl1}. Graphene was grown at a temperature
of 2000$^\circ$C and 1\,atm Ar gas pressure resulting in monolayers
of graphene atomically uniform over more than 1000~$\mu$m$^2$, as
shown by low-energy electron microscopy~\cite{Virojanadara08}. Eight
contacts were produced by depositing 3\,nm of Ti and 100\,nm of Au.
The quadratic sample size of $5 \times 5$~mm$^2$ was achieved by
oxygen plasma etching of all four edges.
Hall measurements indicate that the large area samples are $n$-doped
due to  charge transfer from
SiC~\cite{LaraAvival09,erl1,Bostwick09,SK_TS1a,SK_TS1b,SK_TS2}. The
measured carrier concentration is between $3 \times$10$^{12}$ cm$^{-2}$
and $7 \times$10$^{12}$ cm$^{-2}$, the Fermi energy $E_F$ ranges from  200 to 300~meV and
the mobility is about
1000\,cm$^2$/Vs at room temperature.
%
In these samples, as well as in other large-area samples, the
resistance at room temperature is about 2 to 5~k$\Omega$.

The third epitaxial graphene sample ~\#3-6H, was grown by the
Erlangen group on 6H-SiC(0001) wafers (II-VI Inc.). Graphene growth
was performed using sublimation growth in Ar
atmosphere~\cite{erl1,erl2}. First, polishing damage was removed by
etching the substrate in 1~bar hydrogen at 1550$^\circ$C for 15 min.
Second, graphene was grown by annealing the sample in 1~bar~Ar at a
temperature of 1650$^\circ$C for 15 min. The graphene coverage was
determined by x-ray photoelectron spectroscopy (XPS). The
square-shaped sample size of $4 \times 4$~mm$^2$ was achieved by
mechanical cutting the edges. Both, carrier density and mobility in
the $n$-type sample~\#3-6H are very similar to those of the
Link{\"o}ping samples.

For all epitaxial graphene samples, low-temperature quantum Hall 
measurements reveal the high quality and homogeneity. 



The exfoliated graphene samples (small-area graphene samples~4,~5,
and~6) has been prepared from natural graphite using the
mechanical exfoliation technique~\cite{Bib:Novoselov2004} on an
oxidized silicon wafer. The oxide thickness of $300$\,nm allowed to
locate graphene flakes in an optical microscope and to assess their
thickness. We checked the reliability of this method using Raman
spectroscopy and low-temperature quantum Hall measurements on
similar samples~\cite{jonathan1}. The single layer graphene flakes
obtained by this method were typically $p$-doped by adsorbed
contaminants with carrier concentrations $p \leq 2\times
10^{12}$~cm$^{-2}$. The Fermi energies were $E_F \leq 165$~meV and
the mobilities at room temperature of the order of  $2.5 \times
10^3$~cm$^2$/Vs. The flakes included in this study were all single
layer with the flakes size of the order of 10 to 30~micrometers. The
sample morphology was characterized by atomic force microscopy
measurements under ambient conditions with the microscope in
intermittent contact mode with standard silicon
tips~\cite{jonathan2}. After recording the position of the flakes
with respect to predefined markers, we contacted them by electron
beam lithography and thermal evaporation of 60~nm Pd electrodes. The
resistance of graphene measured between various contacts  was about
1 to 3~k$\Omega$.

\subsection{S2. Laser beam parameters}

In addition to the pulsed THz laser described in the main text,
we also used a continuous-wave (cw) CH$_3$OH laser ($\lambda \, = \, 118\,\mu$m)
with a power of $P \approx $ 20\,mW.
In the experiments applying the CH$_3$OH laser, the cw radiation was
modulated at  
chopper frequencies in the range from $120$ to $600$\,Hz. 
The sign of the signal is defined as a relative phase with respect to the 
lock-in reference signal, which was kept the same for all
measurements. 

Elliptically and, in particular, circularly polarized radiation
has been obtained by transmitting the laser beam, which is initially
linearly polarized along the $y$-direction for the pulsed laser and along the $x$-direction for the cw laser, 
through $\lambda$/4 crystal quartz plates. The resulting polarization state is directly related to the
angle $\varphi$ between the initial linear polarization of the laser light  and the
optical axis of the plate. It is described by the Stokes parameters $S_1, S_2, S_3$~\cite{Stokes}. 
In particular,
the dependence of the circular polarization degree, given by $S_3$, on the angle $\varphi$ in our experimental geometry has the form
\begin{equation} 
\label{Pcirc}
P_{\rm circ} \equiv {S_3}(\varphi) =  \sin{2 \varphi}\:. 
\end{equation}
The parameters $S_1$ and  $S_2$ are given by the bilinear combinations of
the polarization  
vector components, 
\begin{eqnarray} \label{calCS}
&&{S_1}(\varphi) \equiv |e_x|^2 - |e_y|^2 = - \cos^2{2 \varphi} 
\:, \\
&&{S_2}(\varphi) \equiv e_x e_y^* + e_y e_x^* = \frac12 \sin{4
  \varphi} \:. \nonumber  
\end{eqnarray} 
The parameters $S_1$ and $S_2$  describe
the degree of linear polarization in the coordinate axes $x,y$
and in the coordinate frame rotated about an angle of 45$^\circ$,
respectively~\cite{Stokes}.  
Note that radiation is incident along $-z$ axis.
The resulting polarization ellipses for the cw THz laser for some
angles $\varphi$ are sketched on top of Fig.~\ref{figure4}.
Further details on the experimental technique can be found in
Ref.~\cite{GaN2008}. 


\subsection{S3. Photocurrents in small-area samples}

\begin{figure}[t]
\includegraphics[width=0.9\linewidth]{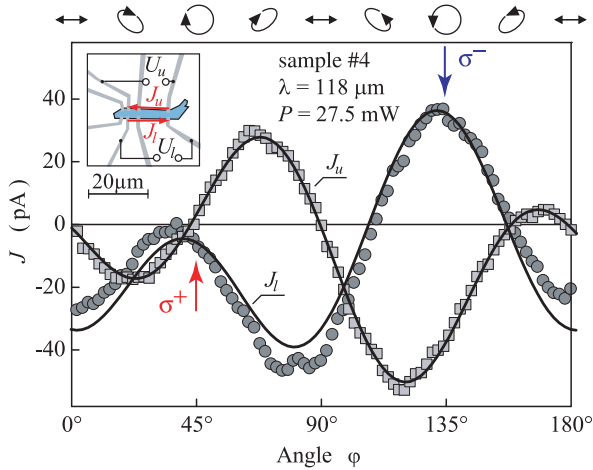}
\caption{
Photocurrent in  small area sample~\#4 as a function of the angle $\varphi$.
Squares and circles show the photocurrent picked up from the contact pairs at the 
upper ($J_u$) and lower ($J_l$) edges of the graphene flake, respectively (see inset).
Full lines show fits to the calculated total current  after Eq.~(\ref{polar_sup}).
The inset demonstrates  the experimental geometry, $U_u$ and $U_l$ are voltages 
recorded by lock-in amplifier. Arrows show current directions for right circularly polarized light.
The signal is  excited by radiation with the wavelength
$\lambda$~=~118~$\mu$m, power $P \approx 20$~mW  and a diameter of the laser spot about $1.5$\,mm.
On top, the polarization ellipses corresponding to various angles $\varphi$ are illustrated.
} \label{figure4}
\end{figure}

Helicity driven photocurrents excited at normal incidence have also
been observed in small-area graphene flakes. 
Examples of the current polarization dependence 
are shown in Fig.~\ref{figure4} for two different pairs of contacts. 
Similar to data obtained 
in large-area samples, it can be well fitted by Eq.~(1) of the main text, which reads
\begin{equation}\label{polar_sup} 
J(\varphi) = J_A P_{\rm{circ}}(\varphi) + J_B S_2(\varphi) + J_C   S_1(\varphi)+ \xi \:.
\end{equation}

While photocurrents are observed in both large-area and small-area
samples,
the analysis of the
edge photocurrents is much easier  in the large-area samples. 
Actually only in the latter type of samples the illumination of a single 
edge by THz radiation could be realized in our experiments. Such
selective excitation has enabled  
the accurate analyses of the edge currents.  
By contrast, in micron-sized exfoliated samples the spot size is much larger 
than the graphene flakes and the
effects of different edges are superimposed. 

%

\subsection{S4. Theory}

Equation~(7) of the main text is obtained by expanding the distribution function $f(\bm{p},x,t)$ in series of the electric field. To first order in the electric field, solution of Eq.~(2) with the boundary condition~(6) has the form
\begin{multline}
f_1 (\bm{p},x) = - \frac{q \tau f'_0}{1-{\rm i}\omega\tau} \left[ \bm{E}_0\cdot\bm{v} - \right.\\
\left. \left( \bm{E}_0\cdot\bm{v} + \frac{\pi}{4} E_{0,x} v \right) \exp \left( - \frac{1-{\rm i}\omega\tau}{v_x \tau}x \right) \Theta (v_x) \right] \:,
\end{multline}
where $f'_0 = d f_0(\varepsilon)/d \varepsilon$. 
The equation for the second-order correction $f_2(\bm{p},x)$ is given by
\begin{equation}\label{equation_f2}
v_x  \frac{\partial f_2(\bm{p},x)}{\partial x}  + 2 q \Re \left[ \bm{E}_0^* \frac{\partial f_1(\bm{p},x)}{\partial \bm{p}} \right] = - \frac{f_2(\bm{p},x)}{\tau} \:,
\end{equation}
which yields 
\begin{multline}\label{f2_dx}
\int_{0}^{\infty} f_2(\bm{p},x) dx = v_x \tau [f_2(\bm{k},0) - f_2(\bm{k},\infty)] \\
-  2q \int_{0}^{\infty} \Re \left[ \bm{E}_0^* \frac{f_1(\bm{p},x)}{d \bm{p}} \right] dx  \:.
\end{multline}
By using Eq.~(4) for the edge electric current and Eq.~(\ref{f2_dx}) we derive
\begin{equation}\label{J_y_2}
J_y = - 8 q^3 \tau^3 \sum_{\bm{p}} \Re \left\{ \frac{v_x v_y \bm{E}_0^*}{1-{\rm i}\omega \tau}  \frac{d [ (\bm{E}_0 \cdot \bm{v}) f'_0 ]}{d\bm{p}} \right.
\end{equation}
\[
\left. + \frac{v_y \bm{E}_0^*}{(1-{\rm i}\omega \tau)^2} \frac{d}{d\bm{p}} \left[
(\bm{E}_0 \cdot \bm{v} + \frac{\pi}{4} E_{0,x} v ) f'_0 \,v_x \right] \right\}  \Theta(v_x)  \:,
\]
%
where the above two contributions to the current stem from the first and second terms on the right-hand side of Eq.~(\ref{f2_dx}), respectively. Finally, taking into account that the electron energy and velocity in graphene 
are given by $\varepsilon_{\bm{p}} = v |\bm{p}|$ and $\bm{v} = v \, \bm{p}/p$, respectively, and assuming that the electron gas is degenerate and $\tau$ is independent of $\varepsilon_{\bm{p}}$, we obtain Eq.~(7) of the main text.
The current Eq.~(7) of the main text is consistent with the point-group symmetry C$_s$ containing the mirror plane 
$(x,z)$. It should be noted that for elliptical polarization, in some experiments the photocurrent 
described by Eq.~(7) is superimposed with an additional  
contribution proportional to $\cos^2{2\varphi}$ [see the third 
term in Eq.~(\ref{polar_sup})].
This term can be attributed to a lowering of the system symmetry to C$_1$ showing the non-equivalence of $y$ and $-y$ directions, e.g.,
due to (i) inhomogeneous photoexcitation, (ii) macroscopic
roughness of the investigated edges,  
(iii)  non-equivalence of contacts, etc. 
The symmetry reduction hinders the edge photogalvanic currents under study and
complicates their analysis.
However, the obstacle can be easily overcome applying circularly polarized radiation, 
like used in the present work. 
Indeed, as addressed above the edge currents driven by circularly polarized light change their
signs upon variation of the radiation helicity. By contrast, other current contributions caused 
by the additional symmetry lowering are insensitive to the radiation helicity and 
vanish for circularly polarized radiation.

%


\end{document}